\documentstyle[ntcs]{article}
\input epsf

\title{\Large\bf Using Dependence Analysis to Support Software \\Architecture 
Understanding}

\author{\large \it{Jianjun Zhao}\\
\normalsize Department of Computer Science and Engineering\\
\normalsize Fukuoka Institute of Technology\\
\normalsize 3-10-1 Wajiro-Higashi, Higashi-ku, Fukuoka 811-02, Japan\\
\normalsize zhao@cs.fit.ac.jp
}

\date{}

\begin{document}

\maketitle

\thispagestyle{empty}

\baselineskip 2.7mm
\abstract
{
Software architecture is receiving increasingly attention as a 
critical design level for software systems. As software architecture 
design resources (in the form of architectural descriptions) 
are going to be accumulated, the development of techniques and tools 
to support architectural understanding, testing, reengineering, 
maintaining, and reusing will become an important issue. 
In this paper we introduce a new dependence analysis technique, named {\it
architectural dependence analysis} to support software architecture 
development. In contrast to traditional dependence analysis, architectural
dependence analysis is designed to operate on an architectural 
description of a software system, rather than the source code of 
a conventional program. Architectural dependence analysis provides 
knowledge of dependences for the high-level architecture of 
a software system, rather than the low-level implementation details 
of a conventional program.}

\vspace{2mm}
\section{Introduction}
\label{sec:intro}
\vspace{2mm}

Software architecture is receiving increasingly attention as a
critical design level for software systems \cite{Shaw96}. The software 
architecture of a system defines its high-level structure, exposing 
its gross organization as a collection of interacting components. 
A well-defined architecture allows an engineer to reason about system 
properties at a high level of abstraction. The importance of software
architecture for practicing software engineers is highlighted by the
ubiquitous use of architectural descriptions in system documentation.

Architectural description languages (ADLs) are formal languages that
can be used to represent the architecture of a software system. 
They focus on the high-level structure of the overall application
rather than the implementation details of any specific source module. 
ADLs are intended to play an important role in the development of
software by composing source modules rather than by composing
individual statements written in conventional programming languages. 
Recently, a number of architectural description languages
have been proposed such as ACME \cite{Garlan97}, Rapide \cite
{Luckham95}, UniCon \cite{Shaw95}, and Wright \cite{Allen96} to
support formally representation and reasoning of software architectures. 
As software architecture design resources (in the form of
architectural descriptions) are going to be accumulated, 
the development of techniques and tools to support understanding, testing,
reengineering, maintaining, and reusing of software architectures 
will become an important issue.  

One promising way to support software architecture development 
is to use dependence
analysis technique. Program dependences are dependence relationships 
holding between program statements in a program that are determined 
by the control flows and data flows in the program. 
Usually, there are two types of 
program dependences in a conventional
program, {\it control dependences} that represent the control
conditions on which the execution of a statement or expression
depends and {\it data dependences} that represent the flow of 
data between statements or expressions. The task to determine a
program's dependences is called {\it program dependence analysis}.
We refer to this kind of dependence analysis 
as {\it traditional dependence analysis} to distinguish it 
from a new form dependence analysis introduced later.

Traditional dependence analysis has been primarily studied 
in the context of conventional programming languages. 
In such languages, it is typically performed 
using {\it program dependence graphs} 
\cite{Cheng97,Horwitz90,Ottenstein84,Zhao96a,Zhao96b}. Traditional 
dependence analysis, though originally proposed for complier
optimization, has also many applications in software engineering 
activities such as program slicing, understanding, debugging, testing,
maintenance and complexity measurement 
\cite{Agrawal93,Bates93,Cheng97,Ottenstein84,Podgurski90,Zhao96a,Zhao96b}. 

Applying dependence analysis to software architectures promises benefit 
for software architecture development at least in two aspects. 
First, architectural understanding and maintenance 
should benefit from dependence analysis. To understand a software 
architecture to make changes during maintenance, a maintainer 
must take into account the many complex dependence
relationships between components and/or connectors in the architecture. 
This makes dependence analysis an essential step to architectural 
level understanding and maintenance. Second, architectural reuse should benefit from 
dependence analysis. While reuse of code is important, reuse of software designs 
and patterns may offer the greater potential for return on investment 
in order to make truly large gains in productivity and quality. 
By analyzing dependences in an architectural description of a software system, 
a system designer can extract reusable architectural descriptions from it, and
reuse them into new system designs for which they are appropriate.

While dependence analysis is useful in software architecture development, 
existing dependence analysis techniques for conventional programming 
languages can not be applied to architectural descriptions 
straightforwardly due to the following reasons. The traditional
definition of dependences only concerned with programs written in 
conventional programming languages which primarily consist of
variables and statements as their basic language elements, 
and dependences are usually defined as dependence relationships
between statements or variables. However, in an architectural
description language, the basic language elements are primarily 
components and connectors, but neither variables nor statements 
as in conventional programming languages. Moreover, in addition 
to definition/use binding relationships, an architectural description 
language topically support more broad and complex relationships
between components and/or connectors such as pipes, event broadcast, 
and client-server protocol. As a result, new types of dependence
relationships in an architectural description must be studied based on
components and connectors.

In this paper we introduce a new dependence analysis technique, 
named {\it architectural dependence analysis} to support software architecture 
development. In contrast to traditional dependence analysis, architectural
dependence analysis is designed to operate on an architectural 
description of a software system, rather than the source code of 
a conventional program. Architectural dependence analysis provides 
knowledge of dependences for the high-level architecture of 
a software system, rather than the low-level implementation details 
of a conventional program.

The purpose of development of architectural dependence analysis 
is quite different from the purpose for development of traditional
dependence analysis. While traditional dependence analysis 
was designed originally for supporting compiler optimization of 
a conventional program, architectural dependence analysis 
was primarily designed for supporting architectural understanding and 
reuse of a large-scale software system. However, just as traditional 
dependence analysis has many other applications in software
engineering activities, we expect that architectural dependence
analysis has also useful in other software architecture development 
activities including architectural testing, reverse engineering, reengineering, 
and complexity measurement. 

The rest of the paper is organized as follows. 
Section \ref{sec:acme} briefly introduces the ACME: an
architectural description language. Section \ref{sec:depen} 
presents a dependence model for software architectures. Section
\ref{sec:use} discusses some applications of the model. 
Concluding remarks are given in Section \ref{sec:final}.

\vspace{2mm}
\section{Architectural Descriptions in ACME}
\label{sec:acme}
\vspace{2mm}

We assume that readers are familiar with the basic concepts of
architectural description languages, and in this paper, we use 
ACME architectural description language \cite{Garlan97} as our target
language to represent software architectures. The selection of the
ACME is based on its potentially wide use because ``it is being 
developed as a joint effort of the software architecture research 
community to provide a common intermediate representation for a wide 
variety of architecture tools.'' \cite{Garlan97}

There are seven design elements in ACME that can be used to represent
software architectures which include components, connectors, systems,
ports, roles, representations, and bindings. Among them, the most
basic elements of architectural description are {\it components}, {\it
connectors}, and {\it systems}. Readers can refer \cite{Garlan97} 
for more details of the language description, and we briefly introduce
these design elements here.

{\it Components} are used to represent the primary computational
elements and data stores of a system. Intuitively, they correspond to
the boxes in box-and-line descriptions of software architectures.
Typical examples of components include clients, servers, filters, 
objects, and databases. Each component has its
interface defined by a set of {\it ports}. A component may
provide multiple interfaces by using different types of ports. 
Each port identifies a point of interaction between the component and its
environment. A port can represent a simple interface such as 
procedure signature, or more complex interfaces, such as a
collection of procedure calls that must be invoked in certain
specified orders, or an event multi-cast interface point.

{\it Connectors} are used to represent interactions between components.
Connectors mediate the communication and coordination activities
between components. Intuitively, they correspond to the lines
in box-and-line descriptions. connectors may represent simple forms 
of interaction, such as pipes, procedure calls, event broadcasts, and
also more complex interactions, such as a client-server protocol 
or a SQL link between a database and an application. Each connector 
has its interface defined by a set of {\it roles}. 
Each role of a connector defines a participant of the interaction 
represented by the connector. Connectors may have two roles such as 
the {\it caller} and {\it callee} roles of an RPC connector, 
the {\it reading} and {\it writing} roles of a pipe, or the {\it
sender} and {\it receiver} roles of a message passing connector, or 
more than two roles such as an even broadcast connector which 
might have a single {\it event-announcer} role and an arbitrary number
of {\it event-receiver} roles.
 
{\it Systems} represent configurations of components and connectors.

\begin{figure}[t]
\hspace*{.7cm}
  \epsfbox{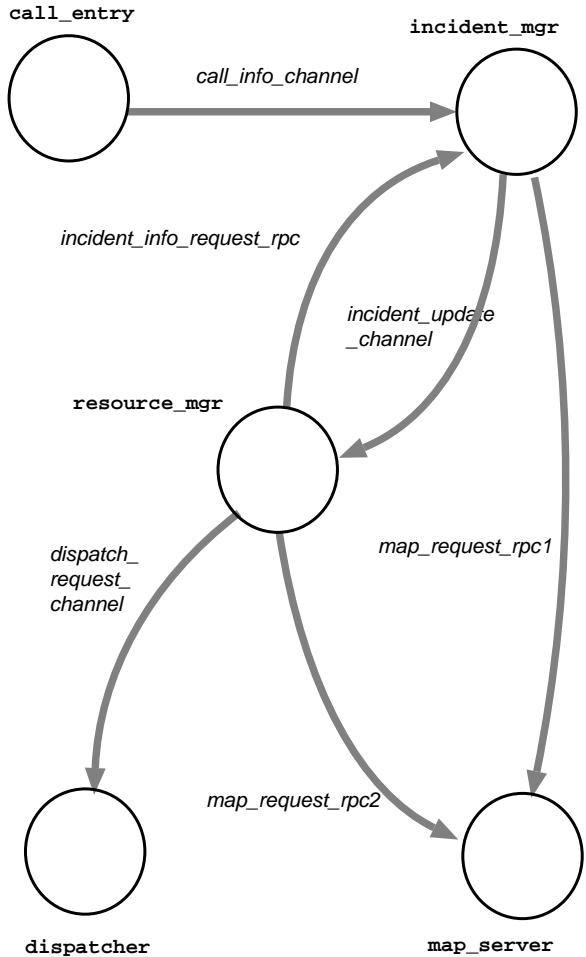}
  \caption{\label{fig:sys} The architecture of the LAS system.}
\end{figure}

Figure \ref{fig:acme} (a) shows the ACME architectural description of 
a simple London Ambulance Service dispatch system (LAS system) which
is taken from \cite{Monroe96}, and Figure 
\ref{fig:sys} shows its architectural representation. 
The architectural representation contains five components which are
connected by six connectors. For example, in the representation, the 
component \verb+call_entry+ and the component \verb+incident_mgr+ is 
connected by the connector \verb+call_info_channel+. Each component 
is declared to have a set of ports, and each connector is declared to 
have a set of roles. For example, a component \verb+incident_mgr+ 
has four ports designed as \verb+map_request+, \verb+incident_info_request+, 
\verb+send_incident_info+, and \verb+receive_call_msg+, and a 
connector \verb+call_info_channel+ has two roles designed as
\verb+from+ and \verb+to+. The topology of the system is declared 
by a set of attachmentses. For example, an attachments 
\verb+incedent_info_path+ represents the connections from calls to 
incident\_manager, incident updates to resource manager, and dispatch 
requests to dispatcher.

In order to provide more information about architectural descriptions, 
ACME also supports annotation of architectural structure with lists of {\it
properties}. Each property has a name, an optional type, and a value,
and each ACME architectural design entity can be annotated. 
For example, in Figure \ref{fig:acme}, the connector
\verb+call_info_channel1+ 
has a set of properties that state the connection type is massage
passing channel and the message flow is from the role \verb+from+ 
to the role \verb+to+. 

In order to focus on the key idea of architectural dependence
analysis, we assume that an ACME architectural description contains 
these basic elements including {\it component} whose interface is
defined by a set of {\it ports}, {\it connector} whose interface is 
defined by a set of {\it roles} and {\it system} whose topology is declared by 
a set of {\it attachmentses} each including a set of attachments. 
{\it Representations} and {\it bindings} will not be considered here, 
and we will consider them in our future work.

\vspace{2mm}
\section{A Dependence Model for Software Architectures}
\label{sec:depen}
\vspace{2mm}

In this section we first introduce three types of
dependences in an architectural description, then present a 
dependence graph for architectural descriptions. 

\subsection{Dependences in Architectural Descriptions}

Traditional dependence analysis has been primarily studied in the
context of conventional programming languages. In such languages,
dependences are usually defined between statements or variables. 
However, in an architectural description language, the basic language
elements are components and connectors, but neither statements nor
variables. 
Moreover, in an architectural description languages, 
the interactions among components and/or
connectors is through their interfaces that are usually defined to be a set 
of ports (for components) and a set of roles (for connectors). 
As a result, it is not enough to define dependences 
just between components and/or connectors in an architectural description. 
In this paper, we define dependences in an
architectural description as dependence relationships between ports
and/or roles of components and/or connectors. In the following, we
present three types of dependences in an architectural description.

\subsubsection{Component-Connector Dependences}

The first type of dependence relationship in an architectural
description is called {\it component-connector dependences} 
which can be used to represent dependence relationships 
between a port of a component and a role of a connector in the
description. Informally, if there is an information flow from a port 
of a component to a role of a connector, then there exists a
component-connector dependence between them. For example, 
in Figure \ref{fig:acme} (a), there is a component-connector 
dependence between the port \verb+receive_incident_info+ of the 
component \verb+resource_mgr+ and the role \verb+to+ of the 
connector \verb+incident_update_channel+ since there is a message 
flow from the role \verb+to+ to the port \verb+receive_incident_info+.

\begin{figure*}[t]
  \begin{center}
     \epsfxsize=\hsize \epsfbox{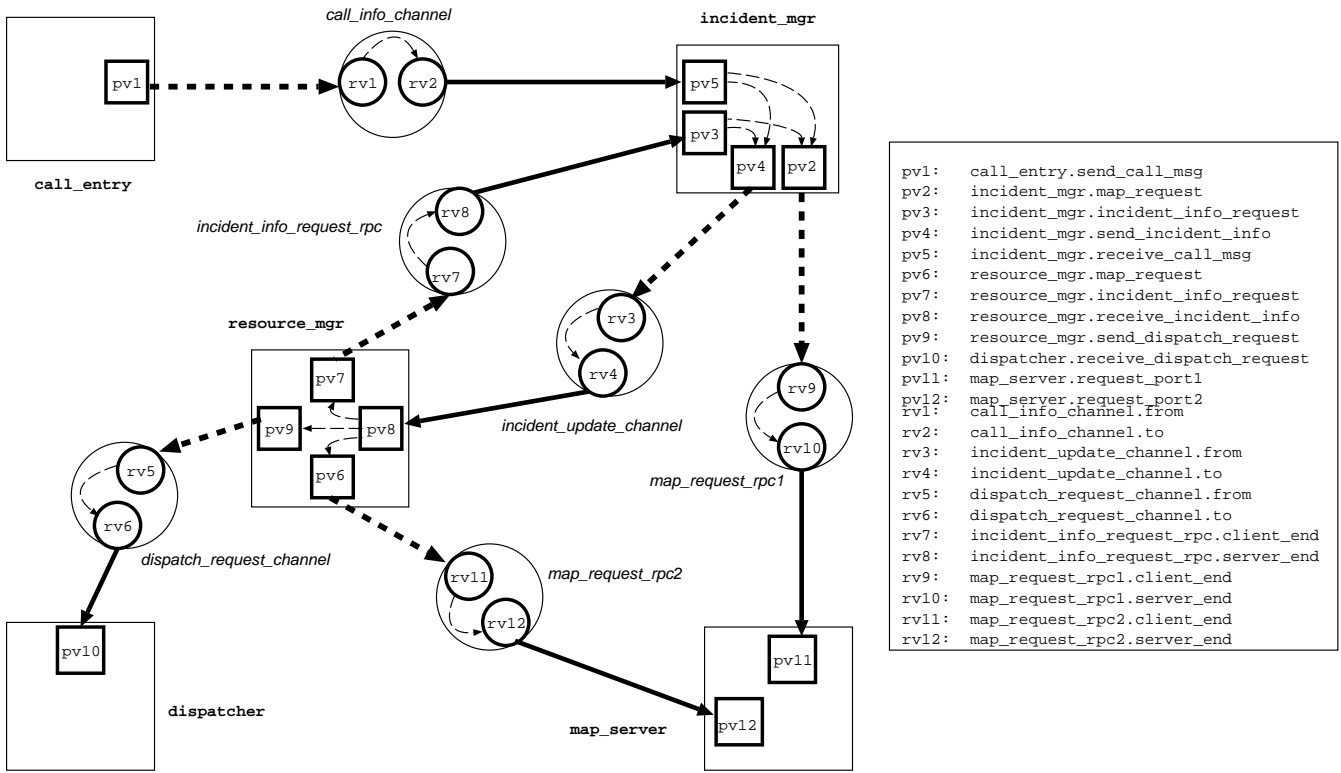}
  \caption{\label{fig:graph} The dependence graph of the architectural
description in Figure {\protect\ref{fig:acme}}.}
  \end{center}
\end{figure*}

\subsubsection{Connector-Component Dependences}

The second type of dependence relationship in an architectural
description is called {\it connector-component dependences} 
which can be used to represent dependence relationships between 
a role of a connector and a port of a component. Informally, 
if there is an information flow from a role of a connector to a port 
of a component, then there exists a connector-component dependence
between them. For example, in Figure \ref{fig:acme} (a), there is a
connector-component dependence between the role
\verb+from+ of the connector \verb+call_info_channel+ and
the port \verb+send_call_msg+ of the component \verb+call_entry+ 
since there is a message flow from the port \verb+send_call_msg+ to
the role \verb+from+.

\subsubsection{Additional Dependences}

The third type of dependence relationships in an architectural
description is called {\it additional dependences} which can be 
used to represent dependence relationships 
between two ports or roles within a component or connector. 
Informally, for a component or connector there are additional
dependences from each port or role as input to other ports or roles as
output. For example, in Figure \ref{fig:acme} (a), there is an additional 
dependence between the roles \verb+client_end+ and
\verb+server_end+ of the connector \verb+map_request_rpc2+ and also an
additional dependence between the ports \verb+map_request+ and
\verb+receive_incident_info+ of the component \verb+resource_msg+.

\subsection{Software Architectural Dependence Graph}

It has been shown that a dependence graph representation such as 
the {\it program dependence graph} (PDG) \cite{Ferrante87,Horwitz90} for
programs written in conventional programming languages, has many
application in software engineering activities since it provides 
a powerful framework for control flow and date flow analysis. 
This motivates us to present a similar representation to explicitly 
represent dependences in an architectural description. In this
section, we present a dependence graph named {\it software 
architectural dependence graph} (SADG for short) for architectural
descriptions to explicitly represent three types of dependences 
in an architectural description introduced above. 
The SADG of an architectural description is an arc-classified digraph 
whose vertices represent the ports of components and the roles of the 
connectors in the description, and arcs represent three types of 
dependence relationships in the description.


\begin{figure*}[t]
  \begin{center}
  \epsfbox{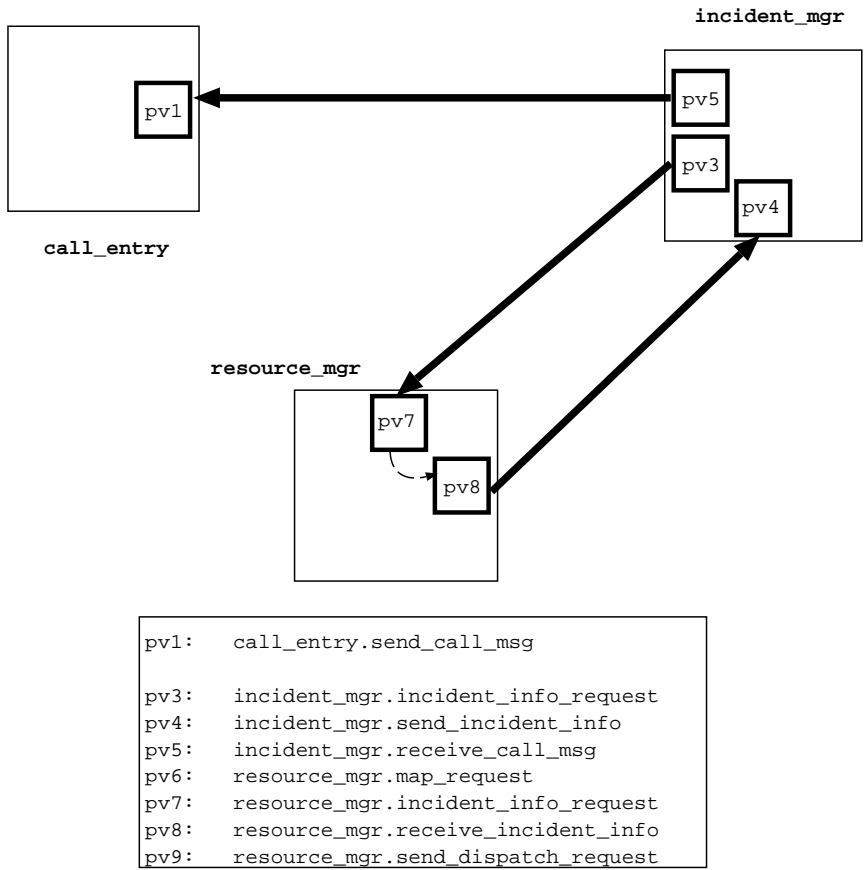}
  \caption{\label{fig:graph.slice} A slice over the ADDG of the 
architectural description in Figure {\protect\ref{fig:acme}}.}
  \end{center}
\end{figure*}

Figure \ref{fig:graph} shows the SADG of 
the architectural description in Figure \ref{fig:acme}. In the figure, 
large squares represent components in the description, and small
squares represent the ports of each component. Each port vertex has
its name described by {\it component\_name.port\_name}. For example, 
$pv8$ (\verb+resource_mgr.receive_incident_info+) is a port vertex 
that represents the port \verb+receive_incident_info+ of the 
component \verb+resource_mgr+. 
Large circles represent connectors in the description, 
and small circles represent the roles of each connector.  Each role
vertex has its name described by {\it connector\_name.role\_name}. 
For example, $rv7$ (\verb+incident_info_request_rpc.client_end+) 
is a role vertex that represents the role 
\verb+client_end+ of the connector \verb+incident_info+ \verb+request+. The
complete description of each vertex is shown in the bottom of the
figure.

Bold arcs represent component-connector dependence arcs that 
connect a port of a component to a role of a corresponding connector. 
Bold dashed arcs represent connector-component dependence arcs that 
connect a role of a connector and a port of a corresponding component. 
Thin dashed arcs represent additional dependence arcs that connect two
ports or roles within a component or connector. For example, 
$(pv8,rv4)$ and $(pv3,rv8)$ are component-connector dependence arcs. 
$(rv5,pv9)$ and $(rv9,pv2)$ are connector-component dependence arcs. 
$(rv2,rv1)$ and $(rv6,rv5)$, and $(pv2,pv5)$ and $(pv7,pv8)$ are
additional dependence arcs.
 
Note that there are some efficient algorithms to compute program
dependences and construct the dependence graph representations 
for programs written in conventional programming languages
\cite{Harrold93,Ottenstein92}. These algorithms can easily be modified
to compute dependences in an architectural description and construct 
the SADG representation as well.

\vspace{2mm}
\section{Applications}
\label{sec:use}
\vspace{2mm}

As dependence graph representations for conventional programming
languages have many applications in software engineering
activities, the dependence model presented in this paper 
should have similar applications in practical development of 
software architectures.

\subsection{Architectural Slicing and Understanding}
\label{subsec:slicing}
 
Program slicing, originally introduced by Weiser \cite{Weiser84}, 
is a decomposition technique which extracts program elements 
related to a particular computation. A {\em program slice} consists of those
parts of a program that may directly or indirectly affect the values 
computed at some program point of interest, referred to as a {\em
slicing criterion}. We refer to this kind of slicing as {\it
traditional slicing}. Traditional slicing has been widely studied 
in the context of traditional programming languages and has many 
applications in software engineering activities such as program 
understanding \cite{DeLucia96}, debugging \cite{Agrawal93}, 
testing \cite{Bates93}, maintenance \cite{Gallagher91} and complexity 
measurement \cite{Ottenstein84}. 

Having SADG as a representation of architectural descriptions, we can 
apply traditional slicing technique to software architectures. We
presented an entirely new form of slicing named {\it architectural
slicing}, to slicing software architectures in order to support 
architectural understanding and reuse \cite{Zhao97}. Architectural 
slicing is designed to operate on the architectural description of a 
software system and can provide knowledge about the high-level 
architecture of a software system. 

Intuitively, an {\it architectural slice} may be viewed as a subset 
of the behavior of a software architectural description, 
similar to the original notion of the traditional static slice.
However, while a traditional slice intends to isolate the behavior
of a specified set of program variables, an architectural slice intends
to isolate the behavior of a specified set of a component's 
ports or a connector's roles.
Given an architectural description $P=(C_{m}, C_{n}, A_{m})$ of a
software system, our goal is to compute a slice 
$S_{p}=(C_{m}', C_{n}', A_{m}')$ that should be a ``subset'' of $P$
that preserves partially the semantics of $P$. 
In \cite{Zhao97}, We use a dependence graph based approach 
to compute an architectural slice, that is based on the SADG of 
the description. Our slicing algorithm contains 
two phases: 

\begin{itemize}
	\item[]{\bf Step 1:} Computing a slice $S_{g}$ over the SADG of an architectural 
description, 
\end{itemize}

Figure \ref{fig:graph.slice} shows a slice over the ADDG in Figure
\ref{fig:graph}. The slice was computed with respect to the slicing
criterion $(\verb+resource_mgr+, V_{c})$ such that 
$V_{c}=\{pv7, pv8\}$. 

\begin{itemize}
	\item[]{\bf Step 2:} Constructing an architectural description slice 
$S_{p}$ from $S_{g}$. 
\end{itemize}

\begin{figure}[t]
  \begin{center}
  \epsfbox{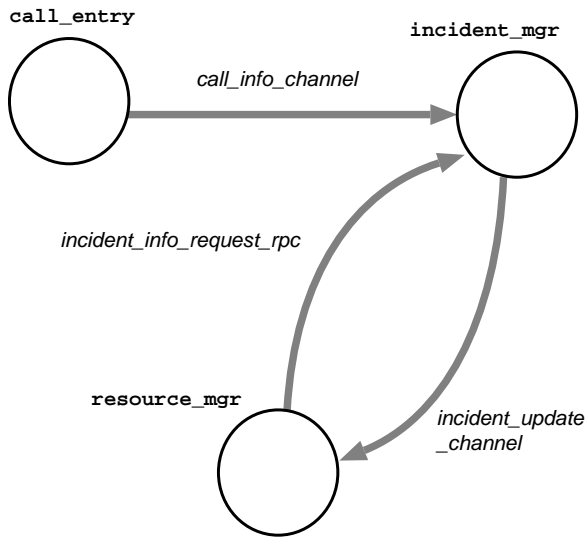}
  \caption{\label{fig:sys-slice} The architectural representation of
the slice in Figure {\protect\ref{fig:acme}} (b).}
  \end{center}
\end{figure}

Figure \ref{fig:acme} (b) shows a slice of the 
ACME description in Figure \ref{fig:acme} (a) with respect to 
the slicing criterion (\verb+resource_mgr+, E) such that 
E=\{\verb+incident_info_request+, \verb+receive_incident_info+\}
is a set of ports of component \verb+resource_mgr+.
The small rectangles represent the parts of description that have been
removed, i.e., sliced away from the original description. 
The slice is obtained from a slice over the ADDG in Figure
\ref{fig:graph.slice} according to the mapping process described above. 
Figure \ref{fig:sys-slice} shows the architectural representation of
the slice in Figure \ref{fig:acme} (b).

In the following, we present a simple example to show how 
architectural slicing can be used to aid architectural 
understanding of a software system.

Consider a simple London Ambulance Service dispatch system (LAS
system) whose ACME description is shown in Figure \ref{fig:acme}
(a). This example is taken from \cite{Monroe96}. 
Suppose a maintainer needs to modify two ports 
\verb+incident_info_request+ and \verb+receive_incident_info+ of the 
component \verb+resource_mgr+ in the architectural description 
in order to satisfy new design requirement, the first thing he/she has
to do is to investigate which components and connectors interact 
with component \verb+resource_mgr+ through these two ports. 
A common way is to manually check 
the source code of the description to find such information. However, 
it is very time-consuming and error-prone even for a small size 
description because there may be complex dependence relations 
between components and/or connectors in the description.
However, if the maintainer has an architectural slicer in hand, 
The work may probably be simplified and automated without the 
disadvantages mentioned above. In such a scenario, he/she only 
needs to invoke the slicer, which takes as input a complete 
architectural description of the system and the set of ports of the 
component \verb+resource_mgr+, i.e., \verb+incident_info_request,
receive_incident_info+ (this is an {\it architectural slicing
criterion}). The slicer then computes an architectural slice 
with respect to the criterion and outputs it to the maintainer. 
Such a slice is a partial description of the original one which
includes those components and connectors that might affect the 
component \verb+resource_mgr+ through ports in the criterion. 
The other parts of the description that might not affect the component
\verb+resource_mgr+ have been removed, i.e., sliced away from the original
description. The maintainer can thus focus his/her attention 
only on the contents included in the slice to investigate the impact
of modification. 

\subsection{Architectural Reuse}
\label{subsec:reuse}

While reuse of code is important, in order to make truly large gains 
in productivity and quality, reuse of software designs and patterns may 
offer the greater potential for return on investment. 
Although there are many researches have been proposed for reuse of
code, little reuse method has been proposed for architectural reuse.
By slicing an architectural description of a software system, 
a system designer can extract reusable architectural descriptions from it, and
reuse them into new system designs for which they are appropriate.

\vspace{2mm}
\section{Concluding Remarks}
\label{sec:final}
\vspace{2mm}

Software architecture is receiving increasingly attention as a 
critical design level for software systems. As software architecture 
design resources (in the form of architectural descriptions) 
are going to be accumulated, the development of techniques and tools 
to support architectural-level understanding, testing, reengineering, 
maintaining, and reusing will become an important issue. 
In this paper we introduce a new dependence analysis technique, named {\it
architectural dependence analysis} to support software architecture 
development. In contrast to traditional dependence analysis, 
architectural dependence analysis is designed 
to operate on an architectural description of a software system, 
rather than the source code of a conventional program. Architectural 
dependence analysis provides knowledge of dependences for the
high-level architecture of a software system, rather than the
low-level implementation details of a conventional program.
In order to perform architectural dependence analysis, we also 
presented the {\it software architectural dependence graph} to 
explicitly represent various types of dependences in an architectural 
description of a software system. While our initial exploration used
ACME as the architectural description language, the concept and
approach of architectural dependence analysis are language-independent.
However, the implementation of an architectural dependence analysis 
tool may differ from one architecture description language to another 
because each language has its own structure and syntax which must be 
handled carefully. 

In architectural description languages, in addition to provide both
a conceptual framework and a concrete syntax for characterizing
software architectures, they also provide tools for parsing, 
displaying, compiling, analyzing, or simulating architectural 
descriptions written in their associated language. However, 
existing language environments provide no tools to support 
architectural understanding, maintenance, testing, and reuse 
from an engineering viewpoint. We believe that a dependence analysis 
tool such as an architectural dependence analyzer introduced in this paper 
should be provided by any ADL as an essential means to support 
software architecture development activities.

As future work, we would like to extend our approach presented in 
this paper to handle other constructs in ACME language such as 
{\it templates} and {\it styles} which were not considered here, and 
also to extend our approach to handle other architecture description 
languages such as UniCon and Wright. Moreover, to demonstrate the 
usefulness of our dependence analysis approach, we are implementing 
an architectural dependence analyzer for ACME architectural
descriptions to support architectural understanding and reuse. 
The next step for us is to perform some experiments to evaluate the 
usefulness of architectural dependence analysis in practical
development of software architectures. 

\vspace{3mm}

\noindent
{\large \bf Acknowledgements}

\vspace{1mm}
The author would like to thank Prof. Jingde Cheng and Prof. Kazuo
Ushijima at Kyushu University for their helpful discussion and 
continuing encouragement.

\vspace*{1.5mm}

\begin{figure*}[t]
  \begin{center}
     \epsfxsize=0.75\hsize \epsfbox{fig5.eps}
  \caption{\label{fig:acme} An architectural description in ACME and a
slice of it.}
  \end{center}
\end{figure*}

\end{document}